\documentclass[]{article}

\usepackage[babel]{csquotes}

\usepackage[style=apa,sortcites=true,sorting=nyt,url=true,backend=biber]{biblatex}

\usepackage{amsmath, amssymb}

\usepackage{subcaption}

\usepackage{url}

\usepackage{breakurl}

\newtheorem{theorem}{Theorem}

\title{An essay on the history of DSGE models\footnote{I am indebted to F. G. Poinsot and F. A. Tohmé for their valuable suggestions.}}
\author{Genaro Martín Damiani\footnote{E-mail: genaro.damiani@uns.edu.ar}\\Departamento de Economía\\Universidad Nacional del Sur\\\\February 2025}
\date{\null}

\begin{document}
	
	\maketitle
	
	\begin{abstract}
		Dynamic Stochastic General Equilibrium (DSGE) models are nowadays a crucial quantitative tool for policy-makers. However, they did not emerge spontaneously. They are built upon previously established ideas in Economics and relatively recent advancements in Mathematics. This essay provides a comprehensive coverage of their history, starting from the pioneering Neoclassical general equilibrium theories and eventually reaching the \textit{New Neoclassical Synthesis} (NNS). In addition, the mathematical tools involved in formulating a DSGE model are thoroughly presented. I argue that this history has a mixed nature rather than an absolutist or relativist one, that the NNS may have emerged due to the complementary nature of New Classical and New Keynesian theories, and that the recent adoption and development of DSGE models by central banks from different countries has entailed a departure from the goal of building a universally valid theory that Economics has always had. The latter means that DSGE modeling has landed \textit{not without loss of generality}. \\
		
	\end{abstract}
	
	\newpage
	\section{Introduction}
	Economics has always been concerned with providing a theoretical framework to evaluate the potential outcomes of different policies. Smith, Ricardo, and Mill made use of deductive but non-formalized schemes to determine in which direction each relevant variable would move in response to policy decisions.\footnote{Of course, they give numerical examples, but these only have illustrative purposes.} Modern economic theory, however, goes beyond just predicting directions; it attempts to predict the magnitudes of the variations in each relevant variable that a certain policy will cause.
	
	Dynamic Stochastic General Equilibrium (DSGE) models are nowadats the standard theoretical framework for quantitative policy-making analyses. First introduced by \textcite{kydland_time_1982}, these sophisticated mathematical formalizations have gained wide acceptance by policy-makers in recent decades. Furthermore, government institutions, especially central banks, have been actively developing them.
	
	It would be inaccurate to claim that these models appeared from outer space. Many of the ideas behind their equations can be traced back to the theories of Classical economists, and most of the mathematical tools on which DSGE models rely are relatively new.\footnote{These ideas have been in continuous evolution, but often at an accelerated pace during economic crises.}
	
	This essay has two purposes. Firstly, it seeks to outline the history of DSGE models in the most comprehensive way possible. Secondly, it is aimed at analyzing some aspects of this history, elucidating why it has been a mixed evolution and not an absolutist or relativist one, the potential reasons behind the convergence towards a \textit{New-Neoclassical Synthesis} (NNS), and the implications of DSGE modeling on the methodological principles of Economics.
	
	The manuscript is organized as follows: Section 2 briefly revisits (focusing on the ideas rather than the models) some of the most relevant Neoclassical, Keynesian, and monetarist general equilibrium analyses; Section 3 formally introduces the mathematical instruments developed during the twentieth century that allowed for a transition into dynamic general equilibrium (DGE) models and eventually into DSGE ones; Section 4 presents the fundamentals of the New Classical and New Keynesian schools of thought; Section 5 provides some insight into the recent history of DSGE models (RBC models, NNS, usage in the field of policy-making, and some of the criticism received after the 2008 crash); and Section 6 contains the final reflections on the history as a whole.\footnote{I believe that the works by the Classics can be overlooked for the sake of conciseness, as their market interconnection ideas materialized in the Neoclassical general equilibrium theory. Although Keynesian and monetarist models are not often classified as general equilibrium ones, as they involve at least two markets in which feedback effects are present, I think they might not be regarded as partial equilibrium models; instead, labeling them as \textit{utterly simplified} general equilibrium models might be more appropriate.}

	\section{Before DGE models}
	
	\subsection{The Neoclassical general equilibrium framework}
	
	The idea of a general equilibrium was originally thought and formalized by \textcite{walras_elements_1874}. He modeled a one-period economy with $n$ competitive, complete, and interconnected markets on which the agents seek to maximize their own welfare. In the basic version (pure exchange economy), Walras assumes the existence of a certain number of commodities whose initial endowments are taken as given. The more complex version (exchange and production economy) forgoes this assumption, making the outputs of each commodity endogenous and dependent on the number of units of each input allocated to its production. In both cases emerges a system of equations for which there is a vector of relative (with respect to a commodity known as the \textit{numeráire}) prices such that the excess of demand in each market is equal to zero, meaning that they are, simultaneously, in equilibrium. An important conclusion that can be found in Walras' work is that the \textit{Classical Dichotomy} holds; i.e., money only affecting nominal variables and not real variables. Other important result is the Walras' Law, which states that, if $n-1$ of the $n$ markets are in equilibrium, the remaining one must also be.
	
	\textcite{pareto_manuel_1909} enhanced the Walrasian general equilibrium framework. He developed and incorporated ordinal utility functions, disaggregated subsystems of equations for each agent, and formulated the notion of \textit{Paretian optimality} so as to provide mathematical proofs of the welfare-maximizing proprieties that a competitive general equilibrium has.\footnote{Pareto's welfare theorems' proofs were not strictly correct. Like Walras, he \textit{ad-hoc} stated that a unique general equilibrium would exist almost every time we had a system with the same number of equations as unknowns. However, I believe they are not to blame for this; they were just less fortunate than \textcite{arrow_existence_1954} or \textcite{mckenzie_equilibrium_1954, mckenzie_existence_1959} in the sense that they did not have any fixed point theorem (presented in the next section) at hand.}
	
	As groundbreaking as general equilibrium analysis may seem in retrospect, initially, it received much less attention than partial equilibrium analysis. It was not until the appearance of the first Keynesian macroeconomic models (commented in the subsequent subsection) that general equilibrium models began to gain popularity.
	
	Apart from Walras and Pareto, the Stockholm School of Economics may have done the most significant works on general equilibrium modeling during this period. \textcite{wicksell_uber_1893} incorporated Bohm-Bawerk's ideas about capital into a general equilibrium framework, building a distribution theory for wages, rent, and interest. Later \parencite*{wicksell_geldzins_1898}, he partially relied on this theory to formulate the notion of the \textit{natural interest rate}; deviations from which are associated with fluctuations in prices that only cease when the market interest rate returns to its natural level. \citeauthor{cassel_theoretische_1919}'s \parencite*{cassel_theoretische_1919} general equilibrium model utilizes a “revealed demand”\footnote{This expression is attributed to \textcite{samuelson_gustav_1993}.} approach rather than a marginal utility one and studies exchange and production in the case of a \textit{uniformly progressing economy}. Perhaps the most widely known general equilibrium model from this school of thought is the H-O model \parencite{heckscher_utrikeshandelns_1919, ohlin_handelns_1924, ohlin_interregional_1933}, which studies how relative factor abundance determines outputs (both in quantities and composition) and trade patterns across countries.
	
	\subsection{Keynesian Economics and general equilibrium}
	
	During the Great Depression, \textcite{keynes_general_1936} attacked the Classical and Neoclassical theoretical frameworks in a rather unconventional way, as they had often been criticized due to their generalistic pretensions. He claimed that they were “applicable to a special case only and not to the general case” (p. 1), a special case whose characteristics “happen not to be those of the economic society in which we actually live” (p. 1). In other words, Keynes criticized the existing theories for their lack of generality. As suggested by the title of his book, he was concerned with developing a \textit{General Theory}, capable of explaining and providing a solution to the ongoing economic crisis.\footnote{\citeauthor{pigou_mr_1936}, who was one of the theorists Keynes criticized most, claimed that “Einstein actually did for Physics what Mr. Keynes believes himself to have done for Economics. He developed a far-reaching generalisation, under which Newton's results can be subsumed as a special case” (1936, p. 115).} To accomplish this, the author introduced the notions of stickiness in prices and wages, money illusion, static expectations, marginal efficiency of capital, liquidity preference, and other nowadays familiar ideas. With these concepts, he built a theory whose main claim is that the economy inevitably oscillates “round an intermediate position appreciably below full employment and appreciably above the minimum employment a decline below which would endanger life” (p. 254), but economic policy has the potential to influence both the mean position and the patterns (i.e., duration and magnitude) of these fluctuations.
	
	Ironically as it may seem, \textcite{hicks_mr_1937} criticized Keynes' work on the basis that it lacked generality. He suggested that “we may call it, if we like, Mr Keynes' special theory” (p. 152). Of course, Hicks' contributions went far beyond a comment. He developed the first version of the IS-LM model, integrating (mostly) Keynesian and Neoclassical ideas into what he referred to as a “Generalized General Theory” (p. 156). In this model, the interaction of the investment goods market and the money market determines national income and interest rates in simultaneous.
	
	There exist other versions of the IS-LM model (varying in complexity and popularity), all of which serve as examples of Keynesian general equilibrium models. Early ones include those from \textcite{harrod_mr_1937} and \textcite{meade_simplified_1937}. The simplified versions still taught in undergraduate courses are based on the works by \textcite{modigliani_liquidity_1944}, \textcite{samuelson_economics_1948, samuelson_economics_1955} and \textcite{hansen_monetary_1949, hansen_guide_1953}.\footnote{These are usually referred to as Neoclassical Synthesis models. However, \citeauthor{patinkin_money_1956}'s \parencite*{patinkin_money_1956} model, which explicitly relies on a microfounded Walrasian general equilibrium framework of four markets (labor, consumption goods, money, and bonds), may be the only one deserving such label.} Some extensions of the basic IS-LM model are the two-period IS-LM model \parencite{modigliani_liquidity_1944}, the AS-AD model \parencite{brownlee_theory_1950}, and the Mundell-Fleming model \parencite{mundell_capital_1963, fleming_domestic_1962}.

	Another well-known (post) Keynesian formulation is the Phillips curve, which posits a long-term trade-off between inflation and unemployment. Originally \parencite{phillips_relation_1958}, it was a partial equilibrium analysis of the labor market that established a connection between the rate of change in the money wage and the unemployment rate. However, as discussed in \textcite{samuelson_analytical_1960}, it was quickly reinterpreted in general equilibrium terms by theories of demand-pull and cost-push, demand-pull, and demand-shift, which hypothesized why the aforementioned relationship between inflation and unemployment might arise.

	A shared feature among these Keynesian general equilibrium models was the absence of \textit{microfoundations}. The relationships between aggregate variables were taken as given rather than explicitly derived from an individual utility-maximizing framework.\footnote{These relationships were frequently simplified by using fixed parameters, such as the marginal propensity to consumption. However, more complex functions could arise, as was often the case with non-transactional money demand.} This turned out to be one of the main reasons of dissatisfaction with Keynesian Economics.
	
	\subsection{Monetarism and general equilibrium models}
	Emerging in the 1950s, the ideas of this school of thought were the first objection to the Keynesian paradigm. One of the innovative elements that made monetarism a formidable challenger was the systematic usage of statistical analysis to test empirically the validity of different theoretical formulations. 
	
	Among their most important contributions were the reinterpretation of \citeauthor{fisher_purchasing_1912}'s \parencite*{fisher_purchasing_1912} quantity equation of money as a money demand curve \parencite{friedman_quantity_1956}, adaptative expectations \parencite{cagan_monetary_1956}, the permanent-income hypothesis \parencite{friedman_theory_1957}, the money multiplier \parencite{brunner_schema_1961, brunner_further_1964}, and the Non-Accelerating Inflation (or natural) Rate of Unemployment \parencite{friedman_role_1968}.
	
	In contrast with Keynesians, monetarists believe that prices and wages are naturally flexible and that any sort of stickiness can only occur due to government intervention. In addition, they think the economy as inherently stable, attributing variations in the quantity of money as the primary cause of (unnecessary) fluctuations in output and employment.
	
	Within the monetarist framework, a long-term trade-off between inflation and employment cannot exist. \textcite{friedman_role_1968} and \textcite{phelps_phillips_1967} argue that high inflation rates can reduce unemployment below its natural rate as long as the observed inflation remains above its expectation; yet in the long term, expected inflation converges to its realization, thus motivating workers to demand higher wages, which ultimately leads back the economy to its natural rate of unemployment. In other words, monetarism claims that the Classical Dichotomy between nominal and real variables holds in the long term.

	While somewhat skeptical about complex mathematical developments, monetarists formulated various extensions of the IS-LM model \parencite{christ_short-run_1967, ott_budget_1965, silber_fiscal_1970, pesek_foundations_1968} and elaborated a few general equilibrium models \parencite{brunner_monetarist_1972, friedman_theoretical_1970, friedman_monetary_1971, silber_fiscal_1970}. Like the Keynesian ones, they lacked solid microfoundations, which later became a source of criticism.
	
	Before ending this section, a common conception between Keynes and Friedman deserves to be remarked: Both valued and sought generality in their theories. When commenting on Marshallian theory, \citeauthor{friedman_methodology_1953} asserted that “it would be highly desirable to have a more general theory...one that would cover...both those cases in which differentiation of product or fewness of numbers makes an essential difference and those in which it does not” (1953, p. 38). Moreover, the quote for which Friedman is most known, even among non-academic people, reflects the generality that his theory attempted to have: Inflation is \textit{always} and \textit{everywhere} a monetary phenomenon.\footnote{Italics are mine.}

	\section{New mathematical methods, new models}
	\subsection{Fixed point theorems}
	Fixed point theorems have been essential for modern economic modeling, since they allow economists to provide a rigorous proof of equilibrium existence when dealing with complex models without a closed solution. So, in this subsection, I present Brouwer's \parencite*{brouwer_uber_1911}, Banach's \parencite*{banach_sur_1922}, Schauder's \parencite*{schauder_fixpunktsatz_1930}, and Kakutani's \parencite*{kakutani_generalization_1941} fixed point theorems. These are the most frequently utilized in general equilibrium, DGE, and DSGE models.\footnote{If interested in proofs, see \textcite{carter_foundations_2001}.}
	
	\begin{theorem}
		\textbf{Brouwer's fixed point theorem}: Let $S$ be a nonempty, compact, and convex subset of a finite-dimensional normed linear space. For every continuous function $f: \ S \to S$, there exists $x \in S$ such that $f(x)=x$. 
	\end{theorem}
	
	\begin{theorem}
		\textbf{Banach's fixed point theorem}: Let $X$ be a complete metric space. For every contraction mapping $f: \ X \to X$, there exists a unique $x \in X$ such that $f(x)=x$.
	\end{theorem}
	\begin{theorem}
		\textbf{Schauder's fixed point theorem}: Let $S$ be a nonempty, compact, and convex subset of a normed linear space. For every continuous function $f: \ S \to S$, there exists $x \in S$ such that $f(x)=x$. 
	\end{theorem}
	\begin{theorem}
		\textbf{Kakutani's Fixed point theorem}: Let $S$ be a nonempty, compact, and convex subset of a finite-dimensional normed linear space. For every closed, convex-valued correspondence $\varphi: \ S \rightrightarrows S$, there exists $x \in S$ such that $\varphi(x)=x$.
	\end{theorem}

	\subsection{Modern optimization theory}
	During the mid-twentieth century, new developments emerged in the field of optimization theory. These works have allowed for dynamic microfounded  (i.e., based on intertemporal maximization of utility or profits) economic models. In this subsection, I shall present Pontryagin's Maximum Principle \parencite{pontryagin_mathematical_1962}, and dynamic programming \parencite{bellman_theory_1954, bellman_dynamic_1957}.\footnote{I will not show Karush-Kuhn-Tucker optimization \parencite{karush_minima_1939, kuhn_nonlinear_1951}, as it is mostly used in finite time horizon models that will not be formally presented in this essay.} The presentation of the former is based on \textcite{intriligator_mathematical_2002}, while the latter follows \textcite{ljungqvist_recursive_2000}.

	\subsubsection{Pontryagin's Maximum Principle}
	Let $t \in [t_0, t_1]$ be a continuum of time, where $t_0$ is the \textit{initial time} and $t_1$ is the \textit{terminal time} (it might be infinite). Let $\mathbf{x_t}=(x_{1t}, x_{2t}, ..., x_{nt}) \in \mathbb{E}^n$ be the vector of \textit{state variables} at any time $t$, living in the euclidean space $E^n$. Let $\mathbf{u_t}=(u_{1t}, u_{2t}, ..., u_{rt}) \in \mathbb{E}^r$ be the vector of \textit{feasible control variables} at any time $t$, living in the convex and compact subset $U$ of the euclidean space $E^r$.
	
	Starting at the \textit{initial state} $\mathbf{x_0}$ and ending at the \textit{terminal state} $\mathbf{x_1}$, the \textit{state trajectory} can be defined as a sequence ${\{\mathbf{x_t}\}}=\{\mathbf{x_t} \in E^n: t_0 \leq t \leq t_1\}$ that is a continuous function of time. The \textit{control trajectory} is a sequence $\{\mathbf{u_t}\}=\{\mathbf{u_t} \in U: t_0 \leq t \leq t_1\}$, which is assumed to be a piece-wise continuous function of time.
	
	The \textit{objective functional} is a mapping from control trajectories to real numbers that takes the form:
	\[ 
	\max_{\{\mathbf{u_t}\}} J=\int_{t_0}^{t_1}I\left(\mathbf{x}, \mathbf{u}\right) \ dt+F\left(\mathbf{x_1}\right)
	\]
	\[\text{s. t.} \]
	\[
	\quad \mathbf{\dot{x}}=f\left(\mathbf{x}, \mathbf{u}\right), \quad \mathbf{x_0} \ \text{given}, \quad \mathbf{x_1} \ \text{given},  \quad {\{\mathbf{u_t}\}} \in U,
	\]
	where $I(\mathbf{x}, \mathbf{u})$ is the \textit{intermediate function}, $F(\mathbf{x_1})$ is the \textit{final function}, and $\mathbf{\dot{x}}=f(\mathbf{x}, \mathbf{u})$ is the \textit{equation of motion}, all of which are assumed to be continuously differentiable.
	
	Note that the equation of motion consists of $n$ differential equations that provide the time rate of change of the state variable. We can define an $n$-dimensional vector of \textit{costate variables} $\mathbf{y}$ (one corresponding to each differential equation) that can be thought of as a set of dynamic Lagrange multipliers. Provided they are nonzero continuous functions of time, the Lagrangian of the problem is:
	\[
	L=\int_{t_0}^{t_1}\left\{I\left(\mathbf{x}, \mathbf{u}\right)+\mathbf{y}\left[f\left(\mathbf{x}, \mathbf{u}\right)\right]-\mathbf{\dot{x}}\right\} \ dt+F\left(\mathbf{x_1}\right).
	\]
	Doing some manipulations, we get:
	\[
	L=\int_{t_0}^{t_1} \left\{H\left(\mathbf{x}, \mathbf{u}, \mathbf{y}\right)+\mathbf{\dot{y}x} \right\} \ dt+F\left(\mathbf{x_1}\right)-\mathbf{y_1x_1+y_0x_0},
	\]
	where $H(\mathbf{x}, \mathbf{u}, \mathbf{y})=I(\mathbf{x}, \mathbf{u})+\mathbf{y}[f(\mathbf{x}, \mathbf{u})]$ is the Hamiltonian.
	
	If the control trajectory changes from $\{\mathbf{u_t}\}$ to $\{\mathbf{u_t} + \Delta \mathbf{u_t}\}$, which implies a change in the state trajectory from $\{\mathbf{x_t}\}$ to $\{\mathbf{x_t} + \Delta \mathbf{x_t}\}$, in an optimum, the following condition must be held:
	\[
	\Delta L= \int_{t_0}^{t_1} \left\{ \frac{\partial H}{\partial \mathbf{u}} \Delta \mathbf{u} + \left(\frac{\partial H}{\partial \mathbf{x}}+ \mathbf{\dot{y}} \right) \Delta \mathbf{x} \right\} \ dt + \left( \frac{\partial F}{\partial \mathbf{x_1}}-\mathbf{y_1} \right) \Delta \mathbf{x_1}=0.
	\]
	Therefore, the necessary conditions for a maximum are:
	\[
	\frac{\partial H}{\partial \mathbf{u}}=\mathbf{0}, \quad \quad \mathbf{\dot{y}}=-\frac{\partial H}{\partial \mathbf{x}}, \quad \quad \mathbf{y_1}=\frac{\partial F}{\partial \mathbf{x_1}}.
	\]
	We can combine the former two. Considering that $\partial H / \partial \mathbf{u}=\partial I/\partial \mathbf{u}+ \mathbf{y} (\partial f/\partial \mathbf{u})$ and deriving with respect to time, the first necessary condition can be re-expressed as:
	\[
	\frac{d}{dt} \left(\frac{\partial I}{\partial \mathbf{u}} \right) +
	\mathbf{\dot{y}} \frac{\partial f}{\partial \mathbf{u}} + \mathbf{y} \frac{d}{dt} \left(\frac{\partial f}{\partial \mathbf{u}} \right)=\mathbf{0} .
	\]
	Replacing with $\partial H / \partial \mathbf{x}=\partial I/\partial \mathbf{x} + \mathbf{y} (\partial f/\partial \mathbf{x})$ in the second necessary condition yields:
	\[
	\mathbf{\dot{y}}= -\left(\frac{\partial I}{\partial \mathbf{x}} + \mathbf{y} \frac{\partial f}{\partial \mathbf{x}} \right)
	\]
	Combining these new expressions results in the \textit{Euler equation}:
	
	\[
	\frac{d}{dt} \left(\frac{\partial I}{\partial \mathbf{u}} \right) 
	-\left(\frac{\partial I}{\partial \mathbf{x}} + \mathbf{y} \frac{\partial f}{\partial \mathbf{x}} \right) \frac{\partial f}{\partial \mathbf{u}} + \mathbf{y} \frac{d}{dt} \left(\frac{\partial f}{\partial \mathbf{u}} \right)=\mathbf{0}.
	\]
	
	Regarding the last of the necessary conditions, it can be written way more elegantly in what is known as the \textit{transversality condition}: 
	\[
	\lim_{t \to t_1} \mathbf{y_t}^\top \mathbf{x_t}=\mathbf{0}.
	\]

	\subsubsection{Dynamic programming}
	I will first show the deterministic version and then the stochastic version as an extension. Let $\mathbf{x_t} \in X$ be the vector of state variables at period $t$, where $X$ is an $m$-dimensional complete metric space. Let $\mathbf{u_t} \in U$ be the $k$-dimensional vector of feasible controls at period $t$, where $U$ is a non-empty, compact, and convex subset of $\mathbb{R}^k$. Define $f(\mathbf{x_t}, \mathbf{u_t})$ as a real-valued one-period return function $r(\mathbf{x_t}, \mathbf{u_t})$ for which continuity, concavity, and boundedness are assumed. Define $\mathbf{x_{t+1}}=g(\mathbf{x_t}, \mathbf{u_t})$ as the \textit{transition equation} (or the \textit{law of motion}) and assume that $g: X \times U \to X$ is continuous.
	
	Starting in $t=0$ from any $\mathbf{x_0} \in X$ given, we have to find a time-invariant \textit{policy function} $h: \mathbf{x_t} \to \mathbf{u_t}$ such that the infinite sequence of controls $\{\mathbf{u_t}\}$ generated by iterating in $\mathbf{u_t}=h(\mathbf{x_t})$ and $\mathbf{x_{t+1}}=g(\mathbf{x_t}, \mathbf{u_t})$ maximizes the infinite sum of discounted returns $\sum_{t=0}^{\infty} \beta^t r(\mathbf{x_t}, \mathbf{u_t})$, where $0 < \beta < 1$.
	
	In order to find $h(\mathbf{x_t})$, it is necessary to define a \textit{value function} $v(\mathbf{x_0})$ that represents the optimal value of the problem when starting from any arbitrary $\mathbf{x_0}\in X$. We can express this function as:
	\[ v\left(\mathbf{x_0}\right) = \max_{\mathbf{\{u_t\}}_{t=0}^\infty} \sum_{t=0}^{n} \beta^t r\left(\mathbf{x_t}, \mathbf{u_t}\right) \quad \text{s. t.} \quad \mathbf{x_{t+1}}=g\left(\mathbf{x_t}, \mathbf{u_t}\right), \quad  \mathbf{x_0} \ \text{given}.
	\]
	The first decision can be considered separately from the future ones, yielding:
	\[
	v\left(\mathbf{x_0}\right)=\max_\mathbf{u_0} \left\{ r\left(\mathbf{x_0}, \mathbf{u_0}\right)+\beta \left[\max_{\mathbf{\{u_t\}}_{t=1}^\infty} \sum_{t=1}^{n} \beta^t r\left(\mathbf{x_t}, \mathbf{u_t}\right) \right] \right\}
	\]
	subject to $\mathbf{x_{t+1}}=g(\mathbf{x_t}, \mathbf{u_t})$, $\mathbf{x_0}$ given. Note that the term in square brackets is equal to the optimal value of the decision problem in $t=1$ starting from $\mathbf{x_1}=g(\mathbf{x_0}, \mathbf{u_0})$. Replacing with $v[g(\mathbf{x_0}, \mathbf{u_0})]$ results in the \textit{Bellman equation}:
	\[
	v\left(\mathbf{x_0}\right)=\max_\mathbf{u_0} \{r\left(\mathbf{x_0},\mathbf{u_0}\right)+\beta v\left[g\left(\mathbf{x_0}, \mathbf{u_0}\right)\right]\}.
	\]
	It can be expressed fancier by dropping the time subscripts:
	\[
	v\left(\mathbf{x}\right)=\max_\mathbf{u} \{r\left(\mathbf{x},\mathbf{u}\right)+\beta v\left[g\left(\mathbf{x}, \mathbf{u} \right) \right] \}.
	\]
	
	This functional equation has a unique solution, which is a concave function. The sketch\footnote{See \textcite{stokey_recursive_1989} for a complete proof.} of the proof goes as follows: The operator
	\[
	Tv=\max_\mathbf{u} \{r\left(\mathbf{x},\mathbf{u}\right)+\beta v\left[g\left(\mathbf{x}, \mathbf{u} \right) \right]\}
	\]
	maps a continuous bounded function $v$ into a continuous bounded function $Tv$. As $T$ satisfies monotonicity and discounting, it meets \citeauthor{blackwell_discounted_1965}'s \parencite*{blackwell_discounted_1965} sufficient conditions to be a contraction mapping. Since $X$ is a complete metric space, by Banach's theorem, $v=Tv$ (another form of expressing the functional equation) has a unique fixed point in $X$. In addition, $T$ maps concave functions into concave functions. Therefore, the solution obtained for the functional equation is unique and it is concave function.
	
	Starting from any bounded and continuous $v_0$, the solution can be approached in the limit ($j \rightarrow \infty)$ by iterating on
	\[
	v_{j+1}\left(\mathbf{x}\right)=\max_\mathbf{u} \{r\left(\mathbf{x,u}\right)+\beta v_j\left[g\left(\mathbf{x}, \mathbf{u} \right) \right]\},
	\]
	which yields an optimal time-invariant policy function $\mathbf{u_t}^*=h(\mathbf{x_t^*})$.
	
	As proven by \textcite{benveniste_differentiability_1979}, $v$ turns out to be differentiable off corners. Let $\mathbf{x}= \mathbf{x_t}$ and $\mathbf{u}= \mathbf{u_t}$. The first-order condition can be written as:
	
	\[
	\frac{\partial r\left(\mathbf{x_t}, \mathbf{u_t}\right)}{\partial \mathbf{u_t}} + \beta v'\left(\mathbf{x_{t+1}}\right) \frac{\partial g\left(\mathbf{x_t}, \mathbf{u_t}\right)}{\partial \mathbf{u_t}} = \mathbf{0}.
	\]
	In addition, we have the envelope condition:
	\[
	v'\left(\mathbf{x_t}\right) = \frac{\partial r\left(\mathbf{x_t}, \mathbf{u_t}\right)}{\partial \mathbf{x_t}} + \beta v'\left(\mathbf{x_{t+1}}\right) \frac{\partial g\left(\mathbf{x_t}, \mathbf{u_t}\right)}{\partial \mathbf{x_t}}.
	\]
	Forward it one period, assume $\partial g(\mathbf{x_t}, \mathbf{u_t})/\partial \mathbf{x_t}=0$, and substitute in the fist-order condition to obtain the Euler equation: 

	\[
	\frac{\partial r\left(\mathbf{x_t}, \mathbf{u_t}\right)}{\partial \mathbf{u_t}} + \beta \frac{\partial r\left(\mathbf{x_{t+1}}, \mathbf{u_{t+1}}\right)}{\partial \mathbf{x_{t+1}}} \frac{\partial g\left(\mathbf{x_t}, \mathbf{u_t}\right)}{\partial \mathbf{u_t}} = \mathbf{0}.
	\]
	
	Now, I will briefly present stochastic dynamic programming. Assume that the transition equation is given by $\mathbf{x_{t+1}}=g(\mathbf{x_t},\mathbf{u_t},\mathbf{e_{t+1}})$, where $\mathbf{e_t}$ is a vector of i.i.d. random variables. As $\mathbf{e_{t+1}}$ is realized at $t+1$ (this is, after $\mathbf{u_t}$ is chosen), any state beyond $\mathbf{x_t}$ is unknown at period $t$. Therefore, the problem becomes:
	\[
	v\left(\mathbf{x_0}\right) = \max_{\mathbf{\{u_t\}}_{t=0}^\infty} \mathbb{E}\left[ \sum_{t=0}^{n} \beta^t r\left(\mathbf{x_t}, \mathbf{u_t}\right) \mid \mathbf{x_0} \right]
	\]
	\[\text{s. t.}
	\]
	\[
	\mathbf{x_{t+1}}=g\left(\mathbf{x_t}, \mathbf{u_t}, \mathbf{e_{t+1}}\right), \quad  \mathbf{x_0} \ \text{given}.
	\]
	The stochastic Bellman equation is:
	\[
	v\left(\mathbf{x}\right)=\max_\mathbf{u} \{r\left(\mathbf{x},\mathbf{u}\right)+\beta \mathbb{E} \left\{v \left[g\left(\mathbf{x}, \mathbf{u}, \mathbf{e}\right) \right] \mid \mathbf{x} \right\} \}.
	\]
	Again, there is a unique concave solution to the functional equation (the previous proof sketch is still valid). In this case, however, we have a \textit{contingency plan} $\mathbf{u_t}=h(\mathbf{x_t})$. It can be approached (starting from any bounded and continuous $v_0$) by iterating on:
	\[
	v_{j+1}\left(\mathbf{x}\right)=\max_\mathbf{u} \{r\left(\mathbf{x,u}\right)+\beta \mathbb{E} \left\{v_j\left[ g\left(\mathbf{x}, \mathbf{u}, \mathbf{e} \right) \right] \mid \mathbf{x} \right\}\}.
	\]
	The first-order condition for this stochastic dynamic programming setting can be expressed as:
	\[
	\frac{\partial r (\mathbf{x_t}, \mathbf{u_t})}{\partial \mathbf{u_t}}  + \beta \mathbb{E} \left[ v' \left(\mathbf{x_{t+1}}\right) \frac{\partial g \left(\mathbf{x_t}, \mathbf{u_t}, \mathbf{e_{t+1}}\right)}{\partial \mathbf{u_t}} \mid \mathbf{x_t} \right] = \mathbf{0}.
	\]
	Our envelope condition is:
	\[
	v' (\mathbf{x_t}) = \frac{\partial r \left(\mathbf{x_t}, \mathbf{u_t}\right)}{\partial \mathbf{x}}  + \beta E \left[ v' \left(\mathbf{x_{t+1}} \right) \frac{\partial g \left(\mathbf{x_t}, \mathbf{u_t}, \mathbf{e_{t+1}} \right)}{\partial \mathbf{x}}  \mid \mathbf{x_t} \right].
	\]
	After assuming $\partial g(\mathbf{x_t}, \mathbf{u_t})/\partial \mathbf{x_t}=0$, combining them yields the stochastic Euler equation:
	\[
	\frac{\partial r \left(\mathbf{x_t}, \mathbf{u_t}\right)}{\partial \mathbf{u}}  + \beta E \left[ \frac{\partial r \left(\mathbf{x_{t+1}}, \mathbf{u_{t+1}}\right)}{\partial \mathbf{x}}  \frac{\partial g \left(\mathbf{x_t}, \mathbf{u_t}, \mathbf{e_{t+1}}\right)}{\partial \mathbf{u}}  \mid \mathbf{x_t} \right] = \mathbf{0}.
	\]

	\subsection{Equilibrium characterization in the absence of a closed solution}
	When dealing with fairly complex models, as a closed solution is nonexistent, we often have to rely on other methods in order to solve them. A frequently utilized approach, especially by the first DSGE models, is to characterize the equilibrium based on its Pareto optimality proprieties, as posited by \textcite{debreu_valuation_1954}. The presentation of this method follows \textcite{stokey_recursive_1989}.
	
	Let $S$ be a normed linear space. There are $I$ consumers, indexed by $i=1,2,..., I$. Consumer i has a set of available commodity consumption choices $X_i \subseteq S$. The preferences of this consumer are represented by a utility function $u_i: X_i \to R$. There are $J$ firms, indexed by $j=1,2,...,J$. Firm $j$ has a set of technological possibilities $Y_j \subseteq S$. This firm evaluates them based on the total profits yielded by each alternative.
	
	An \textit{allocation} $(\mathbf{x}, \mathbf{y})$, where $\mathbf{x}=(x_1, x_2,...,x_I)$ and $\mathbf{y}=(y_1, y_2,...,y_I)$, describes the consumption choice of each consumer and the production choice of each firm. An allocation is \textit{feasible} if $x_i \in X$ for all $i$, $y_j \in Y_j$ for all $j$, and $\sum_{i=1}^{I}x_i - \sum_{j=1}^{J}y_i=0$.
	
	We say that an allocation is \textit{Pareto optimal} if it is feasible and if there is no other feasible allocation such that $u_i(x_i') \geq u_i(x_i) \ \forall i$ and $u_i(x_i') > u_i(x_i)$ for some $i$. A \textit{price system} is defined as a continuous linear functional $\phi: S \to R$. A \textit{competitive equilibrium} $[(\mathbf{x}, \mathbf{y}), \phi]$ is an allocation together with a price system such that $(\mathbf{x}, \mathbf{y})$ is feasible, $x \in X_i$ and $\phi (x) \leq \phi(x_i^0)$ implies $u_i(x) \leq u_i(x_i^0)$ for all $i$, and $y \in Y_j$ implies $\phi (y) \leq \phi(y_j^0)$ for all $j$.
	
	Now, consider these assumptions: 
	\begin{list}{*}{}
		\item $X_i$ is convex for all $i$.
		\item $u_i: X_i \to R$ is continuous. 
		\item For each $i$, if $x, x' \in X_i, u_i(x) > u_i (x'), \ \text{and} \ \theta \in (0,1)$, then we have $u_i [\theta x + (1-\theta)x'>u_i(x')]$. 
		\item The set $Y=\sum_{j=1}^{J}y_j$ is convex. 
		\item Either the set $Y$ has an interior point or $S$ is finite dimensional.
		\item For each $i$ and each $x \in X_i$, there exists a sequence $\{x_n\}$ in $X_i$ converging to $x$ such that $\phi(x_n)<\phi (x)$ for all $n$.
	\end{list}
	If they hold, then every competitive equilibrium is Pareto optimal and vice versa.\footnote{The proof of this statement, which is beyond the scope of this brief presentation, requires an application of the Hahn-Banach theorem (another relatively new mathematical tool).} In Section 5.1, it will be shown how a DSGE model can be solved using this method.

	\section{New Classical and New Keynesian Economics}
	\subsection{The state-of-the-art in the late 1960s}
	Before delving into the New Classical and New Keynesian schools of thought, it is a good idea to present a DGE model from immediately before their emergence. \citeauthor{sidrauski_rational_1967}'s \parencite*{sidrauski_rational_1967} model, built upon Ramsey-Cass-Koopmans's model \parencite{ramsey_mathematical_1928,koopmans_concept_1965,cass_optimum_1965}, offers a very good portrait.
	
	Assume a representative economic unit whose preferences are represented by the strictly concave $C^2$ instantaneous utility function $U_t=U(c_t, m_t)$, where $c_t$ and $m_t$ denote real consumption and real cash balances at time $t$, respectively. Define $m_t=M_t/p_tN_t$, where $M_t$ is the nominal cash balance, $p_t$ is the nominal price of the homogeneous output, and $N_t$ is the number of individuals in the economic unit.
	
	The functional constraints are:
	\[
	a_t=k_t+m_t,
	\]
	\[
	y(k_t)+v_t=c_t+s_t,
	\]
	\[
	s_t=i_t+x_t,
	\]
	\[
	i_t=\dot{k}_t + \left(\delta+n\right) k_t,
	\]
	and
	\[
	x_t=\dot{m}_t+\left(\pi_t + n\right) m_t,
	\]
	where $a_t$ is the total endowment of nonhuman wealth ($a_0$ is assumed to be given), $k_t$ is capital stock, $y(k_t)$ is an homogeneous output, $s_t$ are gross real savings, $i_t$ is gross capital accumulation, $x_t$ is the gross addition to the holdings of real cash balances, $\dot{k}_t$ is the net addition to the capital stock, $\delta$ is the instantaneous rate of capital depreciation, $n$ is the instantaneous rate of growth of the number of individuals in the economic unit, and $\pi$ is the expected rate of changes in prices. 
	
	With some operations, they can be summarized in a stock constraint $a_t=k_t+m_t$ and a flow constraint $\dot{a}_t=y(k_t)+v_t-(\pi_t+n)m_t-(\delta+n)k_t-c_t$. Denote $\rho > 0$ as the subjective rate of time preference of the economic unit, $\lambda_t$ as the Lagrange multiplier attached to the flow constraint, and $q_t$ as the Lagrange multiplier attached to the stock constraint. The Hamiltonian of this problem can be written as:
	\[
	\int_{0}^{\infty} \{U\left(c_t,m_t\right)
	\]
	\[
	+ \ \lambda_t \left[y\left(k_t\right) + v_t-\left(\pi_t+n\right) m_t - \left(\delta+n\right)k_t-c_t \dot{a}_t\right]+q_t\left(a_t-k_t-m_t\right) \} e^{-\rho t}\ dt.
	\]
	We have the following Euler equations:
	\[
	\frac{\partial U\left(c_t, m_t\right)}{\partial c_t}=\lambda_t,
	\]
	\[
	\frac{\partial U\left(c_t, m_t\right)}{\partial m_t}=\lambda_t \left( \pi_t, q_t/\lambda_t + n\right),
	\]
	\[
	\frac{dy}{dk}-\left(\delta+n\right)=q_t/\lambda_t,
	\]
	and
	\[
	\dot{\lambda_t}/\lambda_t=\rho-q_t/\lambda_t.
	\]
	Regarding the transversality condition, it can be expressed as:
	\[
	\lim_{t \to \infty} a_t \lambda_t e^{-\rho t}=0.
	\]
	From these optimality conditions, we can derive demand functions for consumption, real cash balances, and capital, respectively:
	
	\[
	c=c\left(a, \pi, v\right), \quad \quad m=m\left(a, \pi, v\right), \quad \quad k=k\left(a, \pi, v\right).
	\]
	
	Now, consider adaptative expectations of the form $\dot{\pi}=b(\dot{p}/p-\pi)$, where $b>0$. In addition, assume that money supply grows at a constant rate $\theta=\dot{M}/M$ and the population does the same at a constant rate $n=\dot{N}/N$. If each economic unit receives the same amount of net transfers ($v=\theta m$), we can express the aggregate demands for consumption and real cash balances as $\hat{c}=(k, \theta, \pi)$ and $\hat{m}=(k, \theta, \pi)$, respectively.
	
	In an equilibrium growth path, at any time, the money market satisfies the condition
	\[
	M/pN=\hat{m}\left(k, \theta, \pi \right),
	\]
	the rate of change in the capital stock is given by
	\[
	\dot{k}=y\left(k\right) - \hat{c}\left(k, \theta, \pi \right) - \left(\delta+n\right)k,
	\]
	and the expected inflation rate is
	\[
	\dot{\pi}=\frac{\theta - \pi - n - \left[y\left(k\right) - \hat{c}\left(k, \theta, \pi \right)  -\left(\delta +n\right)k \right]}{1+b\left( \partial \dot{m}/ \partial \pi\right) \left( 1/ \dot{m}\right)}.
	\]
	With these equations, we can finally obtain a steady state $\dot{k}=\dot{\pi}=0$ in which the Classical Dichotomy holds:
	\[
	c^*= y\left(k^*\right) - \left(\delta+n\right) k^*, \quad \quad \pi^*=\theta-n.
	\]
	
	As this model shows, immediately before the Rational Expectations revolution, intertemporal microfoundations were already embedded in the body of macroeconomic theory. However, there was a significant flaw in these models: The implicit assumption that agents make optimal decisions over a set of feasible infinite-horizon plans while processing information using a rule of thumb such as adaptative expectations. Fortunately or not, rational expectations were just around the corner.

	\subsection{The New Classicals and the Rational Expectations Revolution}
	This school of thought emerged during the early 1970s, in the context of a supply crisis for which solid explanations were nowhere to be found. Therefore, New Classical economists were mainly concerned with providing plausible explanations for economic cycles. They developed a framework that contemplates intertemporal utility maximization, competitive yet incomplete information markets, and rational expectations to accomplish it.\footnote{In general, incomplete information is modeled considering homogeneous commodity or factor markets composed of different uncommunicated islands \parencite{phelps_new_1969}.}
	
	Rational expectations, albeit not their original development, has been their hallmark. In the words of its creator, John \citeauthor{muth_rational_1961}, this concept means that “the subjective probability distribution of outcomes...tend to be distributed, for the same information set, about...the objective probability distribution of outcomes” (1961, p. 316). This implies that agents make use of the available information in the most efficient way possible in order to forecast the future values of the relevant variables.
	
	In all likelihood, the most emblematic model from this school of thought is Lucas' \parencite*{lucas_expectations_1972, lucas_international_1973, lucas_equilibrium_1975} islands model. Assume an economy with $i=1,2,...,n$ islands. Each representative agent on island $i$ produces a homogeneous commodity $Y_{it}$ using only its labor $L_{it}$, so that $Y_{it}=L_{it}$. The instantaneous utility functions is $U(C_{it}, L_{it})= C_{it}-L_{it}^{1+\gamma}/(1+\gamma)$, where $C_{it}$ is the consumption of agent $i$ at period $t$ and $\gamma$ is the disutility of working. Under the assumption of no investment opportunities, the budget constraint is given by $(P_{it}/P_t)y_{it}=C_{it}$, where $P_{it}$ is the price of commodity $i$ at period $t$ and $P_t$ is the (unknown) price level at period $t$. For the sake of simplicity, certainty-equivalence behavior is assumed. We have to maximize
	\[
	\mathbb{E}_t \sum_{t=0}^{\infty} \beta^t \left[ \left( \frac{P_{it}}{P_t} \right)Y_{it} - \frac{Y_{it}^{1+\gamma}}{1+\gamma}\right].
	\]
	In this simplified version, the first-order condition is:
	\[
	Y_{it}=\left(\frac{P_{it}}{P_t}\right)^{1/\gamma}.
	\]
	Using lowercase letters to denote logarithms of the corresponding uppercase variables, we have the supply function:
	\[
	y_{it}^s=\frac{r_{it}}{\gamma}, 
	\]
	where $r_{it}=p_{it}-p_t$. As said before, $p_t$ is unknown, so $r_{it}$ has to be estimated conditional on the observation of $p_{it}$. Mathematically:
	\[
	y_{it}^s=\frac{1}{\gamma} \mathbb{E}_t \left[r_{it} \mid p_{it} \right].
	\]

	Now, define the aggregate demand (in logarithms) as $y_t^d=m_t-p_t$, where $m_t$ is the money supply at period $t$. It is assumed that the demand for good $i$ at period $t$ is given by
	\[
	y_{it}^d=\left(m_t-p_t\right)+z_{it}-\eta\left(p_{it}-p_t\right), \quad \eta>0,
	\]
	where $z_{it}$ is the good-specific demand shock at period $t$. Provided that $p_t$ and $r_{it}$ are normally and independently distributed, $p_{it}=p_t+r_{it}$ is also normal. Furthermore, $\mathbb{E}_t [r_{it} \mid p_{it} ]$ becomes a linear function of the form:
	\[
	\mathbb{E}_t \left[r_{it} \mid p_{it} \right]= \mathbb{E}_t \left[r_{it} \right] + \frac{\sigma_r^2}{\sigma_r^2+\sigma_p^2} \left(p_{it}- \mathbb{E}_t \left[p_t \right]\right).
	\]
	Since this model is symmetric, $\mathbb{E}_t \left[r_{it} \right]=0$  must be satisfied for all $i=1,2,...,n$. Therefore:
	\[
	\mathbb{E}_t \left[r_{it} \mid p_{it} \right]= \frac{\sigma_r^2}{\sigma_r^2+\sigma_p^2} \left(p_{it}- \mathbb{E}_t \left[p_t \right]\right).
	\]
	The supply curve for agent $i$ at period $t$ becomes:
	\[
	y_{it}^s=\theta \left(p_{it}- \mathbb{E}_t \left[p_t \right]\right), \quad \theta=\frac{1}{\gamma} \frac{\sigma_r^2}{\sigma_r^2+\sigma_p^2}.
	\]
	
	Assuming that both prices and output are equal to their mean in the aggregate, after averaging across producers, we obtain:
	\[
	y_t^s=\theta \left(p_t- \mathbb{E}_t \left[p \right]\right).
	\]
	Recalling that the aggregate demand is given by $y_t^d=m_t-p_t$, in equilibrium:
	\[
	p_t=\frac{1}{1+\theta}m_t+\frac{\theta}{1+\theta} \mathbb{E}_t \left[p_t\right], \quad y_t=\frac{\theta}{1+\theta}m_t-\frac{\theta}{1+\theta} \mathbb{E}_t \left[p_t\right].
	\]
	Taking expectations in the equation of prices leads to the conclusion that $\mathbb{E}_t [p_t]=\mathbb{E}_t [m_t]$. Employing the mathematical artifice $m_t=\mathbb{E}_t [m_t] +(m_t-\mathbb{E}_t [m_t])$, the equilibrium of this economy is described by:
	
	\[
	p_t=\mathbb{E}_t \left[m_t\right]+ \frac{1}{1+\theta} \left(m_t-\mathbb{E}_t \left[m_t\right] \right), \quad y_t=\frac{\theta}{1+\theta} \left(m_t-\mathbb{E}_t \left[m_t\right] \right).
	\]
	The observed component of aggregate demand, $\mathbb{E}_t [m_t]$, has only nominal effects, while its unobserved component, $\mathbb{E}_t [m_t]-m_t$, may have real effects. If there is a higher realization of $m_t$ given its distribution, output will increase in the short term. However, if there is an upward shift in the distribution of $m_t$ but the realization of $m_t-\mathbb{E}_t [m_t]$ is the same, the output level will not be affected. In other words, only unexpected changes in monetary policy can cause fluctuations in output around its natural value, which is equal to 1 (or 0 in logarithms) in this simplified version.
	
	To summarize, in this model, economic cycles are caused by unanticipated changes in the money supply. Due to imperfect information, agents may not tell whether the product they sell has experienced a real or a nominal shift in demand. If they perceive that the former has changed, temporary oscillations in output around its natural trend will occur, at least during the narrow window of time that agents with rational expectations need in order to realize they were wrong. Note that involuntary unemployment cannot exist within this framework (not even during the trough of an economic cycle), as the unemployed agents are those who \textit{choose} not to work at the current price of output.
	
	The works by this group of authors do not end there. A pioneering stochastic partial equilibrium analysis of the labor market that studies the phenomenon of (voluntary) unemployment in further detail can be found in \textcite{lucas_equilibrium_1974}. Regarding fiscal policy, \textcite{barro_are_1974,barro_determination_1979} formalized the concept of Ricardian Equivalence under the assumption of \textit{ultrarrationality}\footnote{Roughly speaking, this notion implies that the households are fully aware of the government's intertemporal budget constraint.}, showing that a fiscal expansion financed by debt has no real effects, even in the short run, as agents are aware that it will lead to heavier tax duties in the future and consequently they save more in the present so as to prepare themselves for the future implications of the policy.
	
	In terms of policy-making, two main developments from the New Classical economists ought to be highlighted.\footnote{Three if we consider the remarkable work by \textcite{sargent_rational_1975}.} The first one is the \citeauthor{lucas_econometric_1976}' \parencite*{lucas_econometric_1976} critique, which posits that non-microfounded models cannot accurately evaluate the outcomes of any policy because they only rely on “superficial” parameters estimated from historical data, failing to account for changes in agents' expectations and behavior that result from any policy. The other one is the concept of a \textit{time inconsistency} introduced by \textcite{kydland_rules_1977}, which suggests that discretionary policy might not achieve its purposes because agents with rational expectations understand that, once they have made their decisions, the policy-maker has an incentive to deviate from its announcements, making this kind of policy non-credible, ineffective, and normatively less desirable than strict policy rules.

	\subsection{The New Keynesian reply}
	
	New Keynesian Economics emerged in the late 1970s after the rational expectations revolution amid macroeconomic instability. While sharing many of the methodological foundations of the New Classicals, it is a New Keynesian belief that monetary and fiscal policy have a significant role as stabilization tools due to wage stickiness, price stickiness, and imperfect competition.
	
	The early works from this school of thought \parencite{fischer_long-term_1977, phelps_stabilizing_1977, taylor_staggered_1979, taylor_aggregate_1980} showed, using partially microfounded models, that monetary policy can have a non-ephemeral influence on the economy even in the presence of rational expectations. 
	
	Fischer's model is great for exposition purposes. Assume that labor contracts are drawn up so as to maintain real wages constant for two periods. More precisely, $w_{t-i}=\mathbb{E}_{t-i}[p_t], \ i=1,2$, where $w_{t-i}$ denotes the nominal wage agreed at period $t-i$ that will prevail at period $t$ and $\mathbb{E}_{t-i}[p_t]$ is the expectation of the log-price level $p_t$ at period $t-i$.
	
	On the supply side, during period $t$, half the firms are operating in the first year of a labor contract signed at $t-1$, while the other half, in the second year of a labor contract signed at $t-2$. The aggregate supply of output is assumed to be given by:
	\[
	y_t^s=\frac{1}{2} \sum_{i=1}^{2}\left(p_t-w_{t-i}\right) + v_t=\frac{1}{2} \sum_{i=1}^{2}\left(p_t-\mathbb{E}_{t-i}\left[p_t\right]\right) + v_t,
	\]
	where $v_t$ is an AR-1 process $v_t=\rho_1 v_{t-1}+\epsilon_t$ (it is assumed that $\epsilon_t$ is an i.i.d. error whose expected value is equal to zero).
	
	Aggregate demand takes the form 
	\[y_t^d=m_t-p_t-z_t,\]
	where $m_t$ is the logarithm of the money supply and $z_t$ is an AR-1 process $z_t=\rho_2 z_{t-1} + \xi_t$ (we again assume that $\xi_t$ is an i.i.d. error with mean zero).
	
	In this model, the rational expectations of $p_t$ at each period can be expressed as follows:
	\[
	\mathbb{E}_{t-2}\left[p_t\right]=\mathbb{E}_{t-2}\left[m_t\right]-\mathbb{E}_{t-2} \left[\left(z_t+v_t\right) \right]
	\]
	and
	\[\mathbb{E}_{t-1} \left[p_t\right]=\frac{2}{3} \mathbb{E}_{t-1} \left[m_t\right]+\frac{1}{3} \mathbb{E}_{t-2}\left[m_t\right] -\frac{1}{3} \mathbb{E}_{t-2} \left[ \left(z_t+v_t\right) \right] - \frac{2}{3} \mathbb{E}_{t-1} \left[ \left(z_t+v_t\right)  \right].
	\]
	
	At this point, assume a money supply rule $m_t=av_{t-1}+bz_{t-1}$, where $a$ and $b$ are fixed by the central bank. All the disturbances are observed at period $t-1$, so there is no difficulty for the public in calculating the money supply at period $t$, thus $\mathbb{E}_{t-1}[m_t]=m_t$. However, at period $t-2$, we have $\mathbb{E}_{t-2}[m_t]=a\rho_1v_{t-2}+b \rho_2 z_{t-2}$.
	
	Having made all these assumptions, we can obtain the output level of this economy:
	\[
	y_t=\frac{1}{2} \left(\epsilon_t+\xi_t\right) \frac{1}{3} \left[\epsilon_{t-1} \left(a+2\rho_1 \right) +\xi_{t-1} \left(b-\rho_2\right) \right] +\rho_1^2 v_{t-2}
	\]
	and its asymptotic variance
	\[
	\sigma_y^2=\sigma_\epsilon^2\left[\frac{1}{4}+\frac{4}{9}\rho_1^2+\frac{\rho_1^4}{1-\rho_1^2}+\frac{a\left(4\rho_1+a\right)}{9}\right] + \sigma_\xi^2 \left[\frac{1}{4}+ \frac{1}{9}\rho_2^2-\frac{b\left(2\rho_2-b\right)}{9}\right].
	\] 
	The conclusion of this model is that, although monetary policy cannot affect the output level in the long term ($\epsilon_{t-1}=\xi_{t-1}=0$), in the short and the medium run, it can affect its \textit{variance}. In the case of this model, the central bank minimizes $\sigma_y^2$ by setting $a=-2\rho_1$ and $b=\rho_2$.
	
	One important critique this model received is that it takes as given the existence of wage stickiness, without explaining how intertemporal optimizing behavior generates it. However, New Keynesian Economics offers three groups of microfounded explanations for this phenomenon: Implicit contracts \parencite{azariadis_implicit_1975, gordon_neo-classical_1976, baily_wages_1974}, efficiency wages \parencite{solow_another_1979, shapiro_equilibrium_1984}, and insiders-outsiders theories \parencite{lindbeck_wage_1986, lindbeck_cooperation_1988}. 
	
	Regarding price stickiness, the existence of menu costs is the main explanation provided by the New Keynesians.\footnote{This idea was originally developed by \textcite{sheshinski_inflation_1977}.} There are two well-known models of staggered prices, both of which are widely used in DSGE modeling.
	
	The first one is \citeauthor{rotemberg_monopolistic_1982}'s \parencite*{rotemberg_monopolistic_1982} model. At any period $t$, there is a price-setting firm facing a trade-off between the cost of changing prices and the cost of deviations from the optimal long-run price $p_t^*$. The intertemporal cost function this firm seeks to minimize can be written as
	\[
	\sum_{s=0}^{\infty} \mathbb{E}_t\left[\alpha \left(p_{t+s}^*-p_{t+s}\right) ^2 + \left(\Delta p_{t+s}\right)^2\right].
	\]
	In $s=0$, by the first-order condition:
	\[
	\Delta p_t= \alpha \left(p_t^*-p_t\right)+ \beta \mathbb{E}_t \Delta p_{t+1} \implies \pi_t=\frac{\alpha}{1+\alpha} \left(p_t^*-p_{t-1}\right) + \frac{\beta}{1+\alpha} \mathbb{E}_t \pi_{t+1}.
	\]
	
	The other model is known as \textit{Calvo pricing} \parencite{calvo_staggered_1983}. At any period $t$, each firm is able to adjust their prices with probability $\rho$, so that $1-\rho$ is the probability of being unable to. Note that the probability of the price $p_t$ remaining equal at period $t+s$ is $(1-\rho)^s$. When firms adjust their prices, they set $p_t$ so as to minimize the cost of deviations from the optimal price $p_t^*$. Their cost function is:
	\[
	\frac{1}{2} \sum_{s=0}^{\infty}\gamma^s \mathbb{E}_t \left[p_t-p_{t+s}^*\right]^2,
	\]
	where $\gamma=\beta(1-\rho)$. The first-order condition of this problem can be written as:
	\[
	p_t=\left(1-\gamma\right)\sum_{s=0}^{\infty}\gamma^s \mathbb{E}_t \left[p_{t+s}^*\right].
	\]
	The general price level is given by:
	\[
	p_t=\rho p_t^*+\left(1-\rho\right)p_{t-1}=\rho \left(1-\gamma\right)\sum_{s=0}^{\infty}\gamma^s \mathbb{E}_t \left[p_{t+s}^*\right] + \left(1-\rho\right)p_{t-1}.
	\]
	Lastly, the inflation rate of this economy is:
	\[
	\pi_t=\rho \left(1-\gamma\right)\sum_{s=0}^{\infty}\gamma^s \mathbb{E}_t \left[p_{t+s}^*-p_{t-1}\right]
	=\frac{\rho \left(1-\gamma\right)}{1-\rho \gamma}\left(p_t^*-p_{t-1}\right)+\frac{\gamma}{1-\rho \gamma} \mathbb{E}_t \left[\pi_{t+1}\right].
	\]
	This pricing mechanism, either in the basic version here presented or with some extensions\footnote{For instance, \textcite{yun_nominal_1996} and \textcite{gali_inflation_1999}.}, is the most common pricing mechanism in DSGE models.

	Imperfect competition models are another significant contribution from this school of thought \parencite{akerlof_can_1985, akerlof_near-rational_1985, svensson_sticky_1986, hart_monopolistic_1979, hart_monopolistic_1982, mankiw_small_1985, mankiw_imperfect_1988, blanchard_monopolistic_1987}. Despite being static and not based on the typical \textcite{dixit_monopolistic_1977} framework, \citeauthor{mankiw_imperfect_1988}'s \parencite*{mankiw_imperfect_1988} model offers a good overview of what these models are, especially in terms of their implications for fiscal policy.\footnote{The implications of imperfect competition for monetary policy will be discussed in Section 5.2, where \citeauthor{ireland_technology_2004}'s \parencite*{ireland_technology_2004} NNS model is presented.}
	
	Assume a representative agent whose utility function can be expressed as $U=\alpha \ln C+(1-\alpha) \ln l$, where $C$ is the consumption of a homogeneous good, $l$ is leisure (the \textit{numeráire} of this economy), and $\alpha \in (0, 1)$. Given the endowment of time $\omega$, the labor income is $\omega-l$. Total net income is given by $\omega-l+\pi-T$, where $\pi$ denotes profits and $T$ is a lump-sum tax. The agent's budget constraint is $PC+l=\omega+\pi-T$, where $P$ indicates the price of the consumption good. 
	
	Although this is not very New Keynesian, Mankiw assumes that the labor market clears. This allows for an application of Walras' Law; all we have to do is find the equilibrium of the commodity market.
	
	On the demand side, solving the consumer's problem
	\[
	\max_C \left\{\alpha \ln C + \left(1-\alpha \right) \ln l\right\}, \quad \text{s. t.} \quad {PC+l=\omega+\pi-T}
	\]
	yields the solution $PC=\alpha(\omega+\pi-T)$. Regarding the government, it has a budget constraint $T=G+W$, where $G$ represents its spending on output and $W$ is its spending on government workers' wages. The total expenditure in the homogeneous commodity is the sum of the consumer's and the government's demands. In other words, $Y=PC+G$, or alternatively, $Y=\alpha (\omega+\pi-T)$.
	
	On the supply side, there are $n$ identical firms. The industry's total output is denoted as $Q$. As total expenditure is taken as given, $Q=Y/P$ is the demand function of the industry. The cost function of each firm is $F+cQ$, where $F$ is overhead labor and $c$ is the labor-to-output ratio. The firms engage in an oligopoly game, from which their profit margin $\mu=(P-c)/P$ is determined. In the short term, as $n$ is fixed, so is $\mu$. 
	
	Total profits are equal to $\pi=PQ-nF-cQ$, or, what is equivalent, $\pi=\mu Y-nF$. By replacing in the aggregate demand equation, we obtain the expression $Y=\alpha(\omega+\mu Y-nF-T)+G$. Solving for $Y$:
	\[
	Y=\frac{\alpha \left( \omega -nF- T\right)+G}{1-\alpha \mu}.
	\]
	From this expression, we can derive the government purchases multiplier and the tax multiplier, respectively:
	\[
	\frac{\partial Y}{\partial G}=\frac{1}{1-\alpha \mu}, \quad \quad \frac{\partial Y}{\partial T}=\frac{-\alpha}{1-\alpha \mu}.
	\]
	Note that imperfect competition ($\mu>0$) is required for a traditional Keynesian multiplier (this is, greater than the unit). In the long term, even without considering the budget constraint, a traditional Keynesian multiplier will not exist because the entry of new firms will make $\mu$ equal to zero.
	
	Combining both multipliers, the balanced budget multiplier ($\Delta G= \Delta T$) can be obtained:
	\[
	\frac{1-\alpha}{1-\alpha \mu}.
	\]
	In the short term, the mechanism behind this multiplier is the following: The initial net increase in expenditure, $(1-\alpha) \Delta G$, raises profits by $ \mu (1-\alpha) \Delta G$, which in turn raises expenditure by $\alpha \mu (1-\alpha) \Delta G$, and so on.

	\section{DSGE Models}
	After revisiting the fundamentals of the New Classicals and the New Keynesians, we can move on to DSGE models. While not presented here, the works by \textcite{lucas_investment_1971} and \textcite{brock_optimal_1972} are examples of stochastic models from the previous decade that preluded what is covered in this section.
	
	\subsection{RBC models}
	Emerging during the early 1980s, in terms of assumptions, DSGE models built by the Real Business Cycles (RBC) school have many things in common with the New Classical DGE models.\footnote{While not revisited here, the works by \textcite{lucas_investment_1971} and \textcite{brock_optimal_1972} are examples of stochastic models from the previous decade.} In consequence, RBC authors are often regarded as a second wave of New Classical economists. However, there are some nuances in their assumptions, as a result of which they arrive to different conclusions. The most important of these nuances is that RBC models incorporate stochastic shocks to technology, accounting for economic fluctuations that are not caused by the government.\footnote{Technological innovations are said to be the main cause behind these shocks, which is indeed an idea \textcite{schumpeter_business_1939} put forward several decades before.}
	
	In the realm of non-monetary RBC models, the most noteworthy are those from \textcite{kydland_time_1982}, \textcite{long_real_1983}, \textcite{hansen_indivisible_1985}, and \textcite{prescott_theory_1986}. Their fundamental premise is that agents cannot fully distinguish between temporary and permanent technology shocks. In response to a positive shock that is perceived as temporary, current investment and working hours tend to increase, leading to higher consumption of goods and leisure in the future when the shock vanishes and productivity returns to its normal level. Therefore, economic cycles are driven by exogenous technological shocks rather than government policy. Indeed, these models do not even include a government! In what follows, I briefly show \citeauthor{prescott_theory_1986}'s \parencite*{prescott_theory_1986} model.
	
	At any period $t$, the output level $Y_t$ is determined by the production function $Y_t=A_t f(K_t, L_t)$, where $Y_t$ represents output, $A_t$ is technology, $K_t$ is capital, and $L_t$ is labor. Constant returns to scale are assumed. Technology follows an AR-1 process given by $A_t=\rho A_{t-1} + \epsilon_t$, where $\epsilon_t$ is an i.i.d. random error with mean zero and $\rho \approx 1$. The prices of capital and labor are equilibrium time-invariant functions of the economy's state $(A_t, K_t$), which means $r_t=r(A_t, K_t)$ and $w_t=w(A_t, K_t)$, respectively. These pricing functions are known to all agents.
	
	This model has a representative household who owns both capital and labor. The household has one unit of time each period, to be allocated between work $L_t$ and leisure $l_t$ such that $L_t+l_t=1$. The household's income is assigned in its totality between consumption $C_t$ and investment $X_t$, subject to the budget constraint $X_t+C_t=w_t L_t+r_t K_t$ and the law of motion of capital $K_{t+1}=(1-\delta)K_t+X_t$, where $\delta$ is the depreciation rate. Assuming a $C^2$ strictly concave instantaneous utility function $U(C_t, l_t)$, the household's optimization problem is
	\[
	\max_{C_t, l_t} \mathbb{E}_t \sum_{t=0}^{\infty} \beta ^t U\left(C_t, l_t\right)
	\]
	
	\[
	X_t+C_t=w_t L_t+r_t K_t, \quad K_{t+1}=\left(1-\delta\right)K_t+X_t, \quad L_t+ l_t=1.
	\]
	The solution consists of two policy functions $C(A_t, K_t)$, $l(A_t, L_t)$, from which and $X(A_t, K_t)$ and $L(A_t,K_t)$ can be derived.
	
	The optimization problem of the representative firm is straightforward. It can be expressed as:
	\[
	\max_{K_t, L_t} =A_t f\left(K_t, L_t\right)-r_t K_t-w_t L_t.
	\]
	Solving it yields two policy functions $K(A_t, K_t)$ and $L(A_t, K_t)$ that allow for the derivation of $Y(A_t, K_t)$.
	
	Even this relatively simplified DSGE model lacks a closed solution.\footnote{It is still possible to obtain an approximated closed solution through a linear-quadratic approximation around the steady state \parencite{kydland_time_1982}, but it requires stronger assumptions about the production function and the utility function.} What can be done, is to characterize the Pareto-optimal equilibrium using Debreu's results. The household's policy functions will be optimal, given the pricing functions and the law of motion of capital. The firm's policy functions will also be optimal, given the pricing functions. There will be spot market clearance, so that $L_t=L(A_t, K_t)$, $K_t=K(A_t, K_t)$, and $X_t(A_t, K_t)+C(A_t, K_t)=Y(A_t, K_t)$. Furthermore, we will have a rational expectations equilibrium, which means that $K_{t+1}=(1-\delta)K_t+X(A_t, K_t)$.
	
	Regarding RBC monetary models, the works by \textcite{king_money_1984}, \textcite{kydland_role_1989}, and \textcite{cooley_inflation_1989} are good references. Although not very useful for policy-making purposes, the incorporation of money in a RBC framework was a prelude of the upcoming extensions that were bound to take place in the following decade and eventually lead to the New Neoclassical Synthesis.
	
	\subsection{The New Neoclassical Synthesis}

	RBC models were just some slight modifications away from becoming powerful instruments for policy-makers. During the 1990s, a new family of DSGE models incorporated in them fiscal policy, market imperfections, and shocks on other relevant variables such as the preferences \parencite{christiano_current_1992, bernanke_financial_1999,clarida_science_1999,kim_monetary_1998, leeper_toward_1994, rotemberg_optimization-based_1997, king_money_1996,benassy_money_1995}. In the first decade of the current century, another family of DSGE models made its appearance, being in essence very similar to the previous, yet having higher mathematical complexity than ever before in order to reach the highest level of accuracy when representing an economy \parencite{adolfson_bayesian_2007,altig_firm-specific_2005,christiano_nominal_2005,iacoviello_house_2005,smets_estimated_2003,smets_shocks_2007}.
	
	As a consequence of these new developments, by the end of the last century, the idea of a New Neoclassical Synthesis (attributed to \textcite{goodfriend_new_1997}) was quickly gaining popularity. Olivier Blanchard \parencite*{blanchard_what_2000} even claimed that the new models “in either the New Keynesian or the New Classical mode (these two labels will soon join others in the trash bin of history of thought) now examine the implications of imperfections, be it in labor, goods or credit markets” (p. 17). But, what is this new consensus all about? \textcite{woodford_convergence_2009} suggests five points in which the vast majority of macroeconomists now agree: Macroeconomics should employ models with rigorous intertemporal general equilibrium foundations, it is desirable to base quantitative policy analysis on econometrically validated structural models, it is important to incorporate expectations as an endogenous variable affected by policy decisions, real disturbances in general (not only technology shocks) are the most important source of economic fluctuations, and monetary policy is effective as a means of inflation control.
	
	To illustrate the fundamentals of this New Neoclassical Synthesis, I present a simplified model elaborated by \textcite{ireland_technology_2004}.\footnote{\textcite{woodford_interest_2003} develops another excellent NNS model.} 
	
	There is a representative household that begins each period $t$ with money $M_{t-1}$ and bonds $B_{t-1}$ (which will mature in the next period). This household supplies $L_t$ units of labor to the intermediate goods-producing firms, earning the nominal wage rate $W_t$. In addition, the household receives nominal profits $D_t$ from these firms and a lump-sum money transfer $T_t$ from the government. The household's income is to be allocated between consumption $C_t$ of the finished good (purchased at the nominal price $P_t$), new bonds valued at $B_t/r_t$ ($r$ is the interest rate between $t$ and $t+1$), and money $M_t$ to be carried into the next period. The household's problem is to maximize
	\[
	\mathbb{E}_t \sum_{t=0}^{\infty} \beta^t \left[z_t \ln C_t+ \ln \left(M_t/P_t\right)- \left(1/\eta \right)L_t^\eta \right]
	\]
	\[
	\text{s. t.}
	\]
	\[
	M_{t-1}+B_{t-1}+T_t+W_tL_t+D_t=P_tC_t+B_t/r_t+M_t,
	\]
	where $\eta \geq 1$. It is assumed that the preferences shock $z_t$ follows an autoregressive process $\ln z_t= \rho_z \ln z_{t-1} + \xi_t$, where $\xi_t$ is a normally distributed i.i.d. error. We have the following solutions:
	\[
	L_t^{\eta-1}=\frac{z_t}{C_t} \frac{W_t}{P_t}
	\]
	and  
	\[
	\frac{z_t}{C_t}=\beta \mathbb{E}_t \left[\frac{z_{t+1}}{C_{t+1}}  \frac{P_t}{P_{t+1}}\right].
	\]
	
	There is a representative finished goods-producing firm operating in a competitive market. It utilizes $Y_{it}$ units of each intermediate good $i\in[0,1]$ (purchased at the nominal price $P_{it}$) to manufacture $Y_t$ units of the finished good. This firm has a constant returns to scale technology:
	\[
	\left[\int_{0}^{1} Y_{it}^{(\theta_t-1)/\theta_t} \ di \right]^{\theta_t/(\theta_t-1)} \geq Y_t,
	\]
	where $\theta_t$ measures the time-varying elasticity of demand for each intermediate good, thus being a cost-push shock. It follows an autorregresive process $\ln \theta_t=(1-\rho_\theta) \ln \theta + \rho_\theta \ln \theta_{t-1} + v_t$, with $\theta >1$ and $0\leq \rho_t<1$, where $v_t$ is a normally distributed i.i.d. error. Profits are maximized by choosing, for all $i \in [0,1]$ and $t=0,1,2...$:
	\[
	Y_{it}=\left(\frac{P_{it}}{P_t}\right)^{-\theta_t}Y_t.
	\]
	In equilibrium, profits are null due to competition. The price of output is then given by:
	\[
	P_t=\left[\int_{0}^{1}P_{it}^{1-\theta_t} \ di\right]^{1/(1-\theta_t)}.
	\]
	
	Regarding the representative intermediate goods-producing firm, it operates in a market of monopolistic competition, hiring $L_{it}$ units of labor to produce $Y_{it}$ units of intermediate good $i$ according to a constants return to scale function
	\[A_t L_{it} \geq Y_{it}.\]
	Technology is assumed to follow a random walk with drift of the form $\ln a_t= \ln A +\ln A_{t-1} + \epsilon_t$, where $a>1$ and $\epsilon_t$ is a normally distributed i.i.d error. We have Rotemberg-type costs of nominal price adjustment:
	\[
	\frac{\phi}{2} \left(\frac{P_{it}}{\pi P_{i t-1}}-1\right)^2 Y_t,
	\]
	where $\phi$ represents the magnitude of the price adjustment cost and $\pi$ is the steady-state inflation rate. This firm has to choose a sequence to maximize its total market value, which is given by:
	\[
	\mathbb{E}_0 \sum_{t=0}^{\infty} \beta^t \left(z_t/C_t\right) \left(D_{it}/P_t\right),
	\]
	where
	\[
	\frac{D_{it}}{P_t}=Y_t\left(\frac{P_{it}}{P_t}\right)^{1-\theta_t}-Y_t
	\left(\frac{P_{it}}{P_t}\right)^{-\theta_t}  \frac{W_t}{P_t} \frac{Y_t}{Z_t}-\frac{\phi}{2} \left(\frac{P_{it}}{\pi P_{i t-1}}-1\right)^2
	\]
	is interpreted as a measure of real profits. We have the following solution: 
	\[
	\left(\theta_t-1\right) \frac{Y_t}{P_t} \left(\frac{P_{it}}{P_t}\right)^{-\theta_t} = 
	\]
	\[
	\theta_t\left(\frac{P_{it}}{P_t}\right)^{-(\theta_t + 1)} \frac{W_t}{P_t} \frac{Y_t}{Z_t} \frac{1}{P_t}- \phi \left[\frac{P_{it}}{\pi P_{i t-1}}-1\right]^2 Y_t \frac{1}{\pi P_{i t-1}}
	\]
	\[ + \beta \phi \mathbb{E}_t \left[\frac{z_{t+1}}{z_t}  \frac{C_t}{C_{t+1}}  \left(\frac{P_{it}}{\pi P_{i t-1}}-1\right) \frac{Y_{t+1}}{\pi P_{it}}  \frac{P_{i t+1}}{P_{it}} \right].
	\]
	
	In a symmetric equilibrium, $Y_{it}=Y_t$, $L_{it}=L_t$, $P_{it}=P_t$, and $D_{it}=D_t$ for all $i \in [0,1]$. In addition, the market-clearing conditions are $M_t=M_{t-1}+T_t$ and $B_t=B_{t-1}$. Solving out for the real wage, hours worked, and real profits, we obtain
	\[
	Y_t=C_t+\frac{\phi}{2}\left(\frac{\pi_t}{\pi}-1\right)^2 Y_t,
	\]
	\[
	\frac{z_t}{C_t}=\beta r_t \mathbb{E}_t\left[\frac{z_{t+1}}{C_{t+1}} \frac{1}{\pi_{t+1}}\right],
	\]
	and
	\[
	\theta_t-1=\theta_t \frac{1}{A_t} \frac{C_t}{z_t} \left(\frac{Y_t}{A_t}\right)^{\eta-1}-\phi \left(\frac{\pi_t}{\pi}\right) \left(\frac{\pi_t}{\pi}-1\right)
	\]
	\[
	+ \beta \phi \mathbb{E}_t \left[\frac{z_{t+1}}{z_t} \frac{C_t}{C_{t+1}}\frac{Y_{t+1}}{Y_t}\frac{\pi_{t+1}}{\pi} \left(\frac{\pi_{t+1}}{\pi}-1\right)\right].
	\]
	
	Define $g_t=Y_t/Y_{t-1}$, $y_t=Y_t/A_t$, $c_t=C_t/A_t$, and $a_t=A_t/A_{t-1}$. In addition, consider $W_t$ as the efficient output level and $x_t=Y_t/W_t$ as the output gap. In the absence of shocks, this economy converges to a steady-state growth path. Along this path, the stationary variables remain constant ($y_t=y$, $c_t=c$, and so on). Using hats to denote the percentage deviation of each variable from its steady-state level (for example, $\hat{y}_t=\ln(y_t/y)$), log-linearization of the model yields:
	\[
	q_t=\rho_z \ln z_{t-1}+ \xi_t,
	\]
	\[
	\hat{e}_t=\rho_\theta \hat{e}_{t-1} + \varepsilon_t,
	\]
	\[
	\hat{a}_t=\epsilon_t,
	\]
	\[
	\hat{x}_t=\mathbb{E}_t\hat{x}_{t+1}-\left(\hat{r}_t-\mathbb{E}_t \hat{\pi}_{t+1}\right)+\left(1-\frac{1}{\eta}\right) \left(1-\rho_z\right) k_t,
	\]
	\[
	\hat{\pi}_t=\beta \mathbb{E}_t \hat{\pi}_{t+1}+\frac{\eta}{\phi} \left(\theta-1\right) \hat{x}_t-\hat{e}_t,
	\]
	and
	\[
	\hat{g}_t=\hat{y}_t-\hat{y}_{t-1}+q_t
	\]
	where $\hat{e}_t=(1/\phi)\hat{\phi}_t=\rho_\theta \hat{e}_{t-1} + \varepsilon_t$ is the transformed cost-push shock (we assume that $\varepsilon_t$ is an i.i.d. normal error). 
	
	We are particularly interested in the versions of the expectational IS curve \parencite{kerr_limits_1996} and the New Keynesian Phillips curve \parencite{roberts_new_1995} this model has. They are, respectively,
	\[
	\hat{x}_t=\mathbb{E}_t\hat{x}_{t+1}-\left(\hat{r}_t-\mathbb{E}_t \hat{\pi}_{t+1}\right)+\left(1-\frac{1}{\eta}\right) \left(1-\rho_z\right) q_t,
	\]
	and
	\[
	\hat{\pi}_t=\beta \mathbb{E}_t \hat{\pi}_{t+1}+\frac{\eta}{\phi} \left(\theta-1\right) \hat{x}_t-\hat{e}_t.
	\]
	These expressions show that a trade-off between inflation and output gap stabilization only arises in the presence of a cost-push shock; this is, shocks to technology or preferences can be perfectly accommodated, at least in theory, by the central bank. Policy-makers can achieve, simultaneously, the stabilization of inflation and output gap by committing to a rule of monetary policy such that the real market interest rate $\hat{r}_t-\mathbb{E}_t \hat{\pi}_{t+1}$ tracks the natural real interest rate $(1-1/\eta)(1-\rho_z)q_t$. Peter Ireland considers a slightly modified \citeauthor{taylor_discretion_1993}'s \parencite*{taylor_discretion_1993} rule that is given by $
	\hat{r}_t-\hat{r}_{t-1}=\theta_\pi\hat{\pi}_t+\theta_g \hat{g}_t + \theta_x \hat{x}_t$, where $\theta_\pi$, $\theta_g$, and $\theta_x$ are response parameters chosen by the central bank so as to minimize some social loss function.

	\subsection{Policy-making}
	The development and use of macroeconomic models for policy-making by central banks is not a recent phenomenon.\footnote{See \textcite{bodkin_history_1991} for a more in-depth discussion.} In the US, this practice began in the 1960s with the FED-MIT-Penn model \parencite{ando_econometric_1969, de_leeuw_federal_1968}, which consisted of a data-driven (if not atheoretical) system of hundreds of equations whose main aim was to predict the effects of alternative economic policies. 
	
	Despite Lucas' critique, replacing these models with microfounded ones took almost two decades. The \textit{FRB/US} model \parencite{brayton_guide_1996} was the first DSGE model developed and used by the FED as a tool for policy-making. By the end of the next decade, most (if not all) central banks developed and implemented one or more DSGE models each \parencite{tovar_dsge_2009}. Given the rationale of this essay, the matter of concern is that the DSGE models developed by the central banks are theoretical constructs designed to encompass the particularities of the modeled economy, being only valid for that economy and under specific circumstances. In other words, the formulation of a universal theory, applicable to all time and space, no longer appears to be the main priority in Macroeconomics.
	
	\subsection{Some critiques on DSGE modeling after the Great Recession}
	Since the 2008 crisis, DSGE models have faced growing criticism because they did not predict it. The more substantial critiques are not those focusing on the outcomes but rather those addressing the methodological mistakes that led to such outcomes. I present the two that may be consider more relevant.\footnote{A more detailed review can be found in \textcite{christiano_dsge_2018}.}
	
	The first one states that the microfoundations underlying DSGE models are not the correct ones, since they do not incorporate the findings of the recent empirical literature regarding the real-world behavior of the agents, as pointed out by renowned economists like Robert \textcite{solow_prepared_2010}, Olivier \textcite{blanchard_dsge_2016}, Joseph \textcite{stiglitz_where_2018}, and others. In response, some recent DSGE models have relaxed the assumption of perfect rationality (in terms of both maximization of intertemporal utility and expectation formation) and conceded a more significant role to the heterogeneity of the agents \parencite{angeletos_quantifying_2018, farhi_monetary_2019, de_grauwe_animal_2015, gabaix_behavioral_2020, massaro_heterogeneous_2013, hommes_behavioral_2023, pfauti_behavioral_2022, bordalo_diagnostic_2018}. A good number of central banks are modifying their DSGE models in this direction, according to a recent survey by \textcite{yagihashi_dsge_2020}.
	
	The second one was done by Paul \textcite{romer_trouble_2016}. He says that the stochastic nature of the DSGE models is a methodologically incorrect approach because it attributes economic fluctuations to imaginary forces, represented by either i.i.d. or Markov error terms. The prominent cause behind fluctuations, i.e., decisions made by agents, remain unaddressed. According to him, the shocks that are present in DSGE models can fall into different categories, such as “a general type of phlogiston that increases the quantity of consumption goods produced by a given output...a troll who makes random changes to the wages...a gremlin who makes random changes to the price of output” (p. 6). Of course, this critique is irreconcilable with DSGE modeling.
	
	\section{Final comments}
	
	As it has been shown, DSGE models are, in essence, not as modern as one might think. Many of their economic foundations had been formulated by earlier theories, some of them dating back more than a hundred years before the appearance of these models. However, what is entirely modern is the mathematical formulation of the ideas, which was rather unfeasible for their original authors due to the lack of the necessary mathematical tools.
	
	Is this history absolutist, relativist, or a mix of both? I believe the latter is the most accurate assessment. Undeniably, the Rational Expectations Revolution started a snowball of knowledge that ultimately led to DSGE models and is still expanding. However, a snowball cannot exist without snow, for if the macroeconomic conditions that prevailed until the early 1970s had not changed, it would have been pointless to develop new theories in order to explain what the old ones could not. Furthermore, if the weather does not allow for its existence, a snowball melts, meaning that continuous changes in the real world (something normal since the 1970s) are required to incentivate new theorizing. Finally, if one attempts to pick up some snow and add it to the snowball without mittens, it will only result in hand ice burns; what I mean by this is that without the context-independent mathematical advancements of the twentieth century, it would have been impossible to represent an economy with a DSGE model. Taking all the factors into account, I believe it is clear that this history has shown (at least up to now) a mixed evolution.
	
	Another question that might come to mind is why there has been a convergence between New Classical and New Keynesian Economics. I hypothesize that the main reason is the complementary nature of both approaches. As we have seen, their models are methodologically very similar; just a few assumptions vary. It might be proper to claim that the New Classicals are more concerned with the long term (where rigidities are non-existent), while the opposite is true for the New Keynesians. Therefore, it seems natural to take the best aspects of both approaches in order to advance towards a more comprehensive theoretical framework. Can a synthesis be established without complementarity? Some cases from other sciences suggest that the answer is negative. In Statistics, no probability interpretation can reconcile frequentists and Bayesians. In Physics, quantum mechanics and relativity theory are still not unified into a unique framework (despite Einstein's attempts!). However, this is merely a hypothesis. There are not enough arguments to categorically support that complementarity is a necessary (much less a sufficient) condition for the merging of two theories.

	Last but not least, DSGE models have landed \textit{not without loss of generality}. In less than 100 years, we have left behind the “generality battle” between Keynes and Hicks, something that is indeed Classical heritage. To a good extent, modern economic theories are meant to be valid only in a specific place and time. This is reflected, for example, in the discussion of the causes behind inflation: Macroeconomics has moved from “inflation is always and everywhere a monetary phenomenon” to a collection of models in which the underlying mechanisms of inflation dynamics are considerably different from one to another.
	
	To conclude this essay, I wish to inform the reader about some of its limitations, the addressing of which remains for future research. It neglects the development of the statistical tools that allowed for estimating DSGE models. The feedback effects between econometric estimation techniques and theoretical modeling of the last few decades have not been considered. The choice of starting point is arbitrary, as the essay could have begun, for example, with the theory of growth. Finally, merely the fundamentals of each school of thought have been covered.

	\newpage
	
	\printbibliography

\end{document}